\pdfoutput=1
\documentclass[twocolumn,10pt]{article}
\usepackage[margin=0.75in]{geometry}
\usepackage{balance}
\usepackage{cite}

\usepackage{graphicx}
\usepackage[font=small,labelfont=bf,textfont=bf,skip=-4pt]{caption}
\usepackage[font=footnotesize,labelfont=bf,textfont=bf,skip=-4pt]{subcaption}
\usepackage{listings}
\usepackage{fancyhdr}
\usepackage[normalem]{ulem}
\usepackage{url}
\usepackage{hyperref}

\newcommand{\ignore}[1]{}

\begin{document}
\sloppy

\title{\bf A New Frontier for Pull-Based Graph Processing}

\author{
    Samuel Grossman \\
    Stanford University \\
    samuelgr@cs.stanford.edu \\
    \and
    Christos Kozyrakis \\
    Stanford University \\
    kozyraki@stanford.edu \\
}

\date{}

\maketitle

\begin{abstract}
The trade-off between pull-based and push-based graph processing engines is well-understood.  On one hand, pull-based engines can achieve higher throughput because their workloads are read-dominant, rather than write-dominant, and can proceed without synchronization between threads.  On the other hand, push-based engines are much better able to take advantage of the \textit{frontier optimization}, which leverages the fact that often only a small subset of the graph needs to be accessed to complete an iteration of a graph processing application.  Hybrid engines attempt to overcome this trade-off by dynamically switching between push and pull, but there are two key disadvantages with this approach.  First, applications must be implemented twice (once for push and once for pull), and second, processing throughput is reduced for iterations that run with push.

We propose a radically different solution: rebuild the frontier optimization entirely such that it is well-suited for a pull-based engine.  In doing so, we remove the only advantage that a push-based engine had over a pull-based engine, making it possible to eliminate the push-based engine entirely.  We introduce \textit{Wedge}, a pull-only graph processing framework that transforms the traditional source-oriented vertex-based frontier into a pull-friendly format called the \textit{Wedge Frontier}.  The transformation itself is expensive even when parallelized, so we introduce two key optimizations to make it practical.  First, we perform the transformation only when the resulting Wedge Frontier is sufficiently sparse.  Second, we coarsen the granularity of the representation of elements in the Wedge Frontier.  These optimizations respectively improve Wedge's performance by up to $5\times$ and $2\times$, enabling it to outperform Grazelle, Ligra, and GraphMat respectively by up to $2.8\times$, $4.9\times$, and $185.5\times$.

\end{abstract}

\section{Introduction}
\label{sec:introduction}

Many application areas, including machine learning, social networking, business intelligence, and bioinformatics, place great importance on solving problems modeled as graphs~\cite{ref_PerformanceStudy,ref_ExperimentalComparison,ref_EvaluationStudy}.  A graph consists of vertices and edges, the former conceptually modelling objects and the latter the relationship between objects.  A graph computation begins by assigning initial values to each vertex.  Vertex values are updated iteratively by transmitting information along the edges as messages to neighbors~\cite{ref_Pregel,ref_PowerGraph}, which are then aggregated locally at each vertex to produce a new value.  The entire process repeats until some application-specific convergence condition is reached.

It is often the case that within one iteration only a subset of vertices receive an updated value that differs from their previous value.  During the following iteration, it is only useful to propagate outbound messages from these \textit{active} vertices. Exploiting this behavior, known as the \textit{frontier optimization}, can significantly reduce the work required to complete an iteration~\cite{ref_Beamer1,ref_Beamer2,ref_Ligra}.  A graph processing framework implements this optimization by tracking which vertices an application updates using a data structure called the \textit{frontier}~\cite{ref_Ligra,ref_Grazelle,ref_GraphMat,ref_Polymer,ref_Pregel}.

Execution of a graph processing application may follow either a \textit{push}~\cite{ref_Grace,ref_GraM,ref_Ligra,ref_Polymer,ref_GraphMat,ref_Graphicionado} or a \textit{pull} processing pattern~\cite{ref_Ligra,ref_Polymer,ref_Gunrock}.  Push-based engines iterate over source vertices and propagate outbound messages. When parallelizing this write-heavy workload, we must use synchronization to avoid conflicting updates to destination vertices. Conversely, pull-based engines iterate over destination vertices and aggregate inbound messages.  Reads are dominant and, since writes occur once per vertex, no synchronization is required when an iteration is parallelized.  Hence, a pull-based engine achieves significantly higher throughput than a push-based engine.  However, the push pattern enables an efficient implementation of the frontier optimization~\cite{ref_Grazelle,ref_PushPull}.  The frontier tracks vertices that are active as sources, which is in alignment with the source orientation of the push pattern.  A push-based engine can simply process the out-edges for the vertices included in the frontier.  Conversely, the pull pattern's destination orientation is out of alignment with the frontier.  A pull-based engine must check every single edge in order to find out if its source vertex is active~\cite{ref_PushPull}.

\begin{figure}[t]
    \begin{center}
        \begin{subfigure}[b]{0.49\linewidth}
            \begin{center}
                \includegraphics{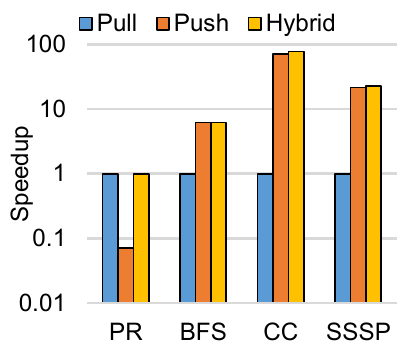}
            \end{center}
            \caption{Overall performance}
            \label{fig:introduction:motivation:performance}
        \end{subfigure}
        \begin{subfigure}[b]{0.49\linewidth}
            \begin{center}
                \includegraphics{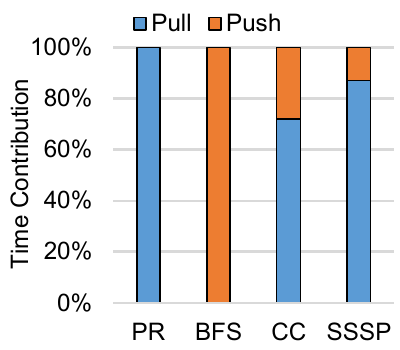}
            \end{center}
            \caption{Hybrid time division}
            \label{fig:introduction:motivation:hybrid}
        \end{subfigure}
    \end{center}
    \caption{Trade-off between push and pull shown through four applications run with Grazelle~\cite{ref_Grazelle} on the \texttt{uk-2007} graph using all cores and sockets of the machine described in \S\ref{sec:evaluation}.  \ref{fig:introduction:motivation:performance} uses pull as the baseline and has a logarithmic vertical axis.}
    \label{fig:introduction:motivation}
  \end{figure}

Figure~\ref{fig:introduction:motivation:performance} quantifies the performance trade-off.  For applications that do not use the frontier optimization, such as PageRank (PR), the higher throughput of a pull engine leads to speedups of $15\times$ over a push engine, a result reflective of the state-of-the-art for both patterns and consistent with Grazelle's published evaluation~\cite{ref_Grazelle}.  On the other hand, the efficient handling of frontiers by the push engine leads to performance gains of up to $82\times$ for frontier-driven applications such as Breadth-First Search (BFS), Connected Components (CC), and Single-Source Shortest Path (SSSP).

Existing graph processing frameworks manage this tradeoff using \textit{hybrid engines} that implement both patterns and dynamically select a push or pull iteration based on the frontier density~\cite{ref_Ligra, ref_Polymer,ref_Grazelle}.  In Figure~\ref{fig:introduction:motivation:performance} the hybrid configuration achieves the best of both worlds and outperforms push-only mode by up to 10\% for the frontier-driven applications shown.  However, there are two key disadvantages of this approach.  First, programmers must write and optimize their graph applications twice, once for the pull and once for the push pattern.  Second, iterations that execute using the push-based engine incur reduced processing throughput. Figure~\ref{fig:introduction:motivation:hybrid} shows the time division between push and pull for the three frontier-based applications we tested when run in hybrid mode.  The time spent on sparse iterations that use the lower-throughput push engine ranges from 10\% to 100\%.  Were these iterations able to use the high-throughput pull engine without sacrificing frontier handling efficiency, the overall performance benefits could be significant.

Rather than continuing to juggle push and pull, we propose eliminating the push pattern entirely.  Our key contribution is \textit{Wedge}, a high-throughput pull-only graph processing framework that implements the frontier optimization efficiently, despite conventional wisdom stating that this is fundamentally impossible~\cite{ref_PushPull}.  Wedge's pull engine continues to produce the traditional source-oriented frontier, but we add a transformation step that converts it to a more pull-friendly format called the \textit{Wedge Frontier}.  The Wedge Frontier is destination-oriented but, rather than tracking active vertices, it tracks individual active edges; simply flipping the frontier to track destination vertices would introduce a potentially unbounded amount of wasted work.

Wedge's transformation step is application-independent, but its overhead can be significant even when parallelized.  Hence, we propose two key optimizations to make it practical.  First, we borrow the key concept of hybrid engines and only transform the traditional frontier into the Wedge Frontier when it is sufficiently empty, thus requiring little work to transform.  Otherwise we simply execute the pull engine on the entire graph without a frontier.  In both cases, Wedge uses the pull engine for processing, improving processing throughput and eliminating the need to implement the application multiple times.  Second, we tweak the granularity of the Wedge Frontier so that one element within it can correspond to multiple edges.  This represents a trade-off between pull engine and transformation step performance: a coarser granularity adds potentially unnecessary work to the pull engine but reduces the transformation step overheads since less elements need to be added to the Wedge Frontier.  The opposite is true of a finer granularity.

Our implementation of Wedge is built on top of \textit{Grazelle}, a state-of-the-art open-source hybrid graph processing framework, resulting in a new pull-only version.  Wedge's two key optimization strategies respectively improve its performance by up to $5\times$ and $2\times$, enabling it to outperform existing graph processing frameworks Grazelle, Ligra, and GraphMat respectively by up to $2.8\times$, $4.9\times$, and $185.5\times$.

Wedge is publicly available on GitHub.  It can be accessed at \url{https://github.com/stanford-mast/Wedge}.

\section{Background}
\label{sec:background}

Our focus is in-memory graph processing on a single server machine, although the techniques we propose with Wedge are not conceptually restricted to this setup.  A modern server can house several terabytes of DRAM and many real-world problems can comfortably fit within this capacity~\cite{ref_LAWS,ref_SNAP}.  Processing graphs in memory on a single machine leads to significantly higher performance than can be achieved using distributed~\cite{ref_Pregel,ref_GPS, ref_Mizan, ref_Pegasus, ref_GraphX, ref_PowerLyra, ref_3DPart,ref_GraM,ref_GraphLab} or out-of-core~\cite{ref_GridGraph,ref_X-Stream,ref_GraphChi,ref_FlashGraph,ref_Chaos} approaches.

\begin{figure}[t]
    \begin{center}
        \begin{subfigure}[b]{0.49\linewidth}
            \begin{center}
                \includegraphics{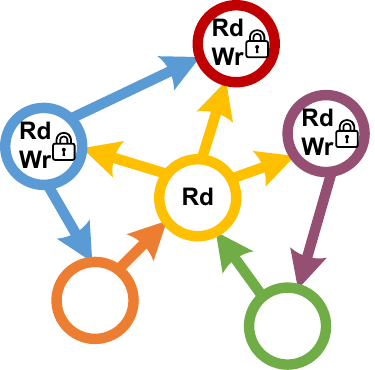}
            \end{center}
            \caption{Push}
            \label{fig:background:pushpull:push}
        \end{subfigure}
        \begin{subfigure}[b]{0.49\linewidth}
            \begin{center}
                \includegraphics{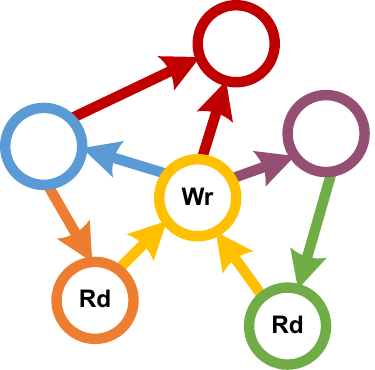}
            \end{center}
            \caption{Pull}
            \label{fig:background:pushpull:pull}
        \end{subfigure}
    \end{center}
    \caption{Illustration of push-based and pull-based graph processing.  Colored edges are grouped with similarly-colored vertices.  Annotations show memory operations for processing the yellow vertex.}
    \label{fig:background:pushpull}
\end{figure}

A graph processing application proceeds in two phases.  In the first phase, messages are exchanged along edges and aggregated at each destination vertex.  In the second phase, local computations are performed on each vertex to produce its updated value based on the aggregation of incoming messages.  An application alternates between these two phases iteratively until some convergence condition is reached.  If a graph engine follows the Bulk-Synchronous Parallel model~\cite{ref_BSP} and completes processing of all vertices in one phase before moving onto the next, it is said to be \textit{synchronous}, otherwise it is \textit{asynchronous}~\cite{ref_PowerGraph}.  Existing work has found that there is no clear winner between these two types of graph engines~\cite{ref_PowerGraph,ref_Hsync}.  We focus on synchronous processing because of its relative simplicity.

\subsection{Push vs. Pull}
\label{sec:background:pushpull}

The fundamental unit of work in the first phase of graph processing is a single edge: a message is propagated from the vertex at the source to the vertex at the destination.  While edges can be processed in any order, it is common to group them by source or destination vertex.  The former grouping produces a \textit{push} processing pattern (Figure~\ref{fig:background:pushpull:push}), whereby a vertex is read once and its outgoing message is aggregated at its outbound neighbors. If we parallelize the first phase, this write-heavy workload requires synchronization because threads may conflict as they update destination vertices.  Conversely, the latter grouping produces a \textit{pull} processing pattern (Figure~\ref{fig:background:pushpull:pull}), in which inbound messages are read, and the result of the aggregation is written a single time to each destination vertex.  This workload is dominated by reads and requires no synchronization because each vertex receives exactly one write.  In terms of programmability the two patterns are equivalent: any application can be represented using either~\cite{ref_PushPull}.  However, the read-heavy and unsynchronized pattern of a pull-based engine leads to significantly higher processing throughput than a push-based engine, as reflected in the PageRank results of Figure~\ref{fig:introduction:motivation:performance}.

Despite its higher throughput, a pull engine severely disadvantage compared to a push engine
in terms of its ability to implement the \textit{frontier optimization}.  This important optimization exploits the common application behavior that only a subset of the graph may need to be processed during each iteration.  Subset size varies per iteration and can be as small as a single-digit number of edges~\cite{ref_Ligra}.  More specifically, an application will add a vertex to the set of active vertices---called the \textit{frontier}---during a particular iteration if its value is updated during that same iteration and so should be transmitted outbound to its neighbors during the next iteration.  Adding a vertex to the frontier during one iteration means that message propagation must occur along all its out-edges during the next.  Out-edges from vertices not in the frontier can be skipped.

A push engine iterates over source vertices and is therefore properly aligned with the frontier.  It can implement the frontier optimization by iterating over the vertices present in the frontier and processing the out-edges associated with each.  In contrast, a pull engine iterates over destination vertices and, as a result is, out of alignment with the frontier.  It must examine each individual incoming edge before it knows the source vertex and is able to check its frontier membership.  Whereas a push engine consults the frontier before accessing any edges, the pull engine must access edges before checking the frontier and must therefore unconditionally scan through all the edges in the graph~\cite{ref_PushPull}.  Despite its lower throughput, the frontier optimization is so effective that, per Figure~\ref{fig:introduction:motivation:performance}, a push engine can achieve a speedup of up to $82\times$ over a pull engine on frontier-driven applications.

\begin{figure}[t]
    \begin{center}
        \includegraphics{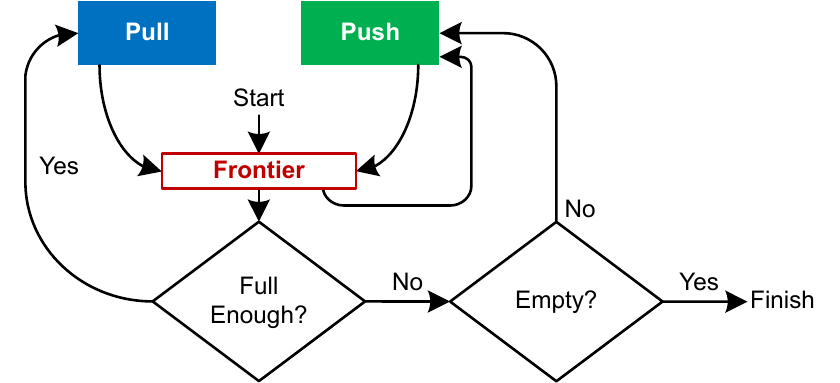}
    \end{center}
    \caption{Hybrid graph processing framework.}
    \label{fig:background:hybrid}
\end{figure}

Previous work has proposed hybrid graph processing frameworks as a means of overcoming the trade-off between push and pull~\cite{ref_Ligra,ref_Polymer,ref_Grazelle}, an approach that, per Figure~\ref{fig:introduction:motivation:performance}, is able to achieve the best of both worlds.  A block diagram illustrating the top-level control flow of a hybrid framework is shown in Figure~\ref{fig:background:hybrid}.  When the frontier is almost full, both push and pull would propagate messages along a very high fraction of the edges in the graph, so the deciding factor is throughput and therefore pull is selected.  Conversely, when the frontier is sufficiently empty, enough work would be saved by exploiting the frontier optimization that choosing push is worthwhile despite its lower throughput.  In determining which engine to use, a hybrid framework computes the sum of the out-degrees of all of the vertices in the frontier and compares the result to a pre-determined threshold, typically a certain fraction of the total number of edges in the graph.  The exact value to use is often determined experimentally~\cite{ref_Ligra,ref_Grazelle}, and the decision of push or pull is made per iteration.  While effective from a performance perspective, the disadvantage of a hybrid engine is that the application writer must implement the application twice (once for push and once for pull).

\subsection{Graph Representation}
\label{sec:background:representation}

Implementations of both push-based and pull-based engines are highly dependent upon the data structures used to represent vertex values, edge information (topology plus optional weight values), and the frontier.  Since every vertex has a value, it is common  to represent vertex values with an array indexed by vertex identifier~\cite{ref_Ligra,ref_Polymer,ref_X-Stream,ref_Grazelle,ref_GraphMat,ref_GraM}.

Edge information is often represented using a two-level data structure known as Compressed-Sparse.  Each instance of this data structure can either represent out-edges (Compressed-Sparse-Row, or CSR) or in-edges (Compressed-Sparse-Column, or CSC). A hybrid graph processing framework would create one instance of each type for the push (CSR) and pull (CSC) engines~\cite{ref_Ligra}.  The top level in Compressed-Sparse, the \textit{vertex index}, contains one element per vertex and indicates that vertex's starting position within the bottom level \textit{edge array}.  The latter contains one element per edge, each element identifying the vertex at the other end of the edge.  Edge weights can be encoded directly into the edge array or placed into a parallel array.

Vector-Sparse, introduced with Grazelle~\cite{ref_Grazelle}, is a modified form of Compressed-Sparse with two key functional differences.  First, edges in the edge array are packed into vectors of up to four edges per vector, all of which correspond to the same vertex in the vertex index.  Second, each such vector encodes the identity of the corresponding vertex in the vertex index.  As a result, it is possible to process the entire graph by streaming through the edge vector array without ever consulting the vertex index.

The frontier is a data structure that tracks active vertices.  In existing work it is often implemented densely as a bit-mask with one bit per vertex~\cite{ref_Ligra,ref_Polymer,ref_Grazelle,ref_GraphMat} or sparsely as a list of active vertices~\cite{ref_Ligra,ref_Polymer}.  Two instances of this data structure typically exist: one is consumed during an iteration of the application while the other is simultaneously being produced for the next iteration.  Because a vertex is defined as being ``active'' if it has an updated value to propagate out-bound to its neighbors, the frontier optimization is fundamentally source-oriented.

\section{Wedge}
\label{sec:wedge}

Wedge is a pull-only graph processing engine that distinguishes itself from all prior work by including an efficient pull-oriented version of the frontier optimization.  This allows Wedge to overcome the trade-off between the push and pull patterns and to eliminate the need for a push engine altogether.  Wedge's frontier optimization design is application- and framework-independent.  When integrated into a specific graph processing framework there is no impact on its programming model, and no changes are required to graph applications other than to remove the now-unnecessary push-based version.

We built the new pull-oriented frontier optimization based on the three key design requirements described in \S\ref{sec:wedge:design}. We summarize the high-level operation of Wedge in \S\ref{sec:wedge:system} and present in more detail its two key components, the transformation step and the new frontier representation, in \S\ref{sec:wedge:frontier}.  Finally, in \S\ref{sec:wedge:optimizations} we describe the techniques we use to make Wedge's frontier optimization design practical.

\subsection{Design Requirements}
\label{sec:wedge:design}

\begin{figure}[t]
    \begin{center}
        \begin{subfigure}[b]{0.49\linewidth}
            \begin{center}
                \includegraphics{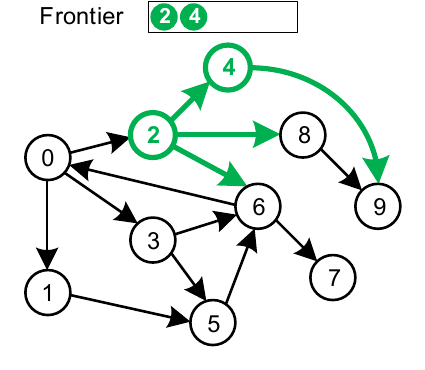}
            \end{center}
            \caption{Source orientation}
            \label{fig:wedge:frontier:frontiersource}
        \end{subfigure}
        \begin{subfigure}[b]{0.49\linewidth}
            \begin{center}
                \includegraphics{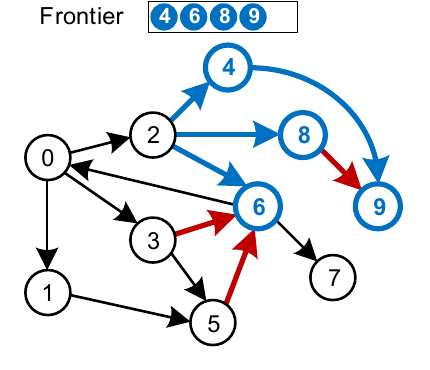}
            \end{center}
            \caption{Destination orientation}
            \label{fig:wedge:frontier:frontierdest}
        \end{subfigure}
    \end{center}
    \caption{Vertex-based frontier following both source and destination orientations.  The highlighted set of 4 active edges is the same in both, but the destination orientation adds 3 superfluous edges.}
    \label{fig:wedge:frontier}
\end{figure}

Sections \ref{sec:introduction} and \ref{sec:background} explained why a pull-based graph engine cannot exploit a traditional source-oriented vertex-based frontier.  The key issue is the misalignment between the frontier's source orientation and the pull engine's destination orientation, which forces a pull engine to scan unconditionally through the entire graph.  Given that the conventionally-built frontier optimization does not work for pull engines, our first goal is to establish the requirements for a version of the frontier optimization that does.

\textbf{Requirement 1: Insertion into the frontier must be vertex-based and source-oriented.}
We begin by observing that the difficulty a pull engine faces does not lie in \textit{inserting} vertices into the frontier.  Both push-based and pull-based engines ultimately compute updated values for vertices, and both are equally capable of knowing which vertices they are updating at the time they produce these updates.  Since the traditional source-oriented vertex-based method of insertion is not an issue, and since any other method of insertion would incur additional processing overhead in the pull engine, our first requirement is that this style of insertion be preserved.

\textbf{Requirement 2: Traversal of the frontier must be destination-oriented.}
Our ultimate goal is to arrive at a frontier that can be traversed using a destination orientation.  Figure~\ref{fig:wedge:frontier} illustrates precisely what this means on an example graph.  Suppose an application adds vertices 2 and 4 to the frontier, marking them active as source vertices.  The active subset of the graph consists of the four highlighted edges in Figure~\ref{fig:wedge:frontier:frontiersource}, all of which are out-edges of vertices 2 and 4.  The equivalent destination-oriented frontier, shown in Figure~\ref{fig:wedge:frontier:frontierdest}, contains the destination vertices of each edge in the active subset of the graph, namely vertices 4, 6, 8, and 9; the out-edges highlighted in Figure~\ref{fig:wedge:frontier:frontiersource} are highlighted as in-edges in Figure~\ref{fig:wedge:frontier:frontierdest}.  If we construct a destination-oriented frontier containing these four vertices, a pull engine would be properly aligned with it, could iterate efficiently over its member vertices, and in so doing would propagate messages along all of the edges in the active subset of the graph.

\textbf{Requirement 3: Filtering out inactive edges within each destination vertex must be supported.}
While switching from a source vertex orientation to a destination vertex orientation solves the problem of misalignment between pull engine and frontier, doing this alone is insufficient.  Highlighted in Figure~\ref{fig:wedge:frontier:frontierdest} are seven edges in total, four of which (blue edges) comprise the active subset, and three of which (red edges) are extra edges that the pull engine would process unnecessarily.  These useless edges are present because activating a vertex as a destination means processing all of its in-edges, while only some of them are part of the active subset.  In Figure~\ref{fig:wedge:frontier:frontierdest} only three additional edges are highlighted, representing an overhead of 75\%.  However, if vertex 6 hypothetically had an in-degree of 1 million, the overhead would quickly dominate.

\subsection{Wedge Overview}
\label{sec:wedge:system}

\begin{figure}[t]
    \begin{center}
        \includegraphics{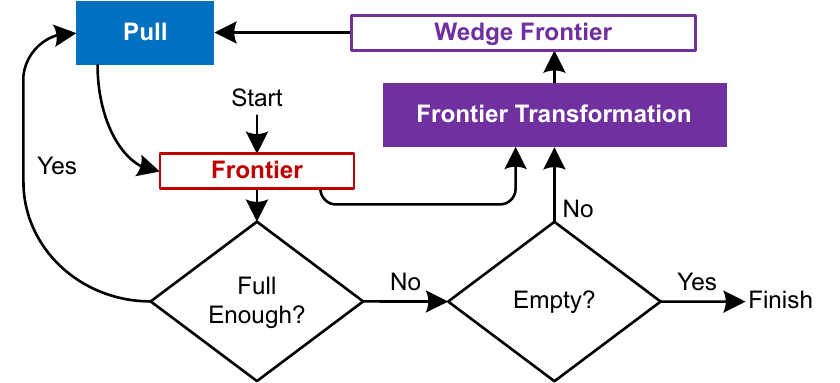}
    \end{center}
    \caption{Wedge graph processing framework.}
    \label{fig:wedge:system}
\end{figure}

Figure~\ref{fig:wedge:system} shows a top-level block diagram of Wedge, our proposed pull-only graph processing framework.  Compared to a hybrid framework (Figure~\ref{fig:background:hybrid}), the push engine has been replaced with a frontier transformation step that consumes the traditional source-oriented vertex-based frontier and produces the \textit{Wedge Frontier} (\S\ref{sec:wedge:frontier}), which contains the same information as the traditional frontier but is formatted such that a pull-based engine can traverse it efficiently.  The pull engine continues to produce the traditional source-oriented vertex-based frontier as output, just as it did in the hybrid framework, meaning that this system design follows Requirement 1.  The transformation step itself is application-independent, as it is simply converting from one representation format to another.  All of the application-specific logic remains encapsulated within the pull engine, meaning that Wedge introduces no impact on programmability, and using it requires no code changes to applications other than to remove the now-obsolete push-based version.

\subsection{Wedge Frontier}
\label{sec:wedge:frontier}

\begin{figure}[t]
    \begin{center}
        \includegraphics{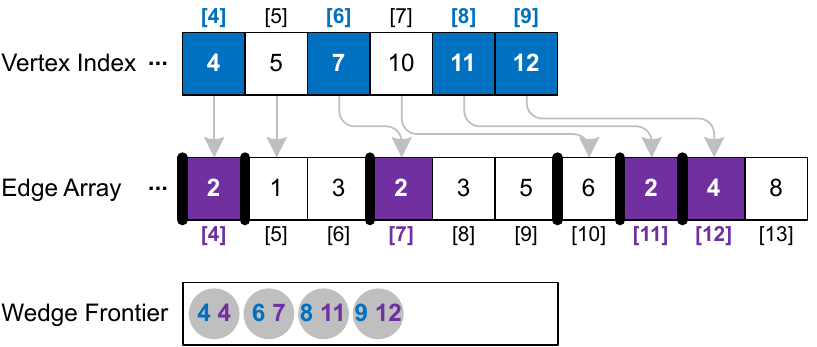}
    \end{center}
    \caption{Wedge Frontier for the example in Figure~\ref{fig:wedge:frontier}, shown using Compressed-Sparse-Column.  Blue-highlighted vertices are the same as in Figure~\ref{fig:wedge:frontier:frontierdest}.  The active subset consists of the purple-highlighted edges, represented in the Wedge Frontier by position in the vertex index and edge array.}
    \label{fig:wedge:frontierwedge}
\end{figure}

The primary difference between the Wedge Frontier and the traditional frontier is that the Wedge Frontier is edge-based rather than vertex-based.  The traditional frontier identifies the active subset of the graph by explicitly identifying which vertices are members of the frontier such that the active subset implicitly consists of all of the in-edges or out-edges of those member vertices.  Conversely, the Wedge Frontier directly identifies the edges that comprise the active subset.  More concretely, whereas values in the traditional frontier are vertex identifiers, values in the Wedge Frontier identify edges by their positions within the edge topology data structure.  The exact meaning of each value stored in the Wedge Frontier is therefore dependent on the data structure selected for representing edge topology.  As many frameworks use Compressed-Sparse, Figure ~\ref{fig:wedge:frontierwedge} shows the content of the Wedge Frontier for the example in Figure~\ref{fig:wedge:frontier} assuming the use of CSC to represent in-edges.  Each element in the Wedge Frontier identifies position in both the vertex index and edge array.  A Vector-Sparse version (\S\ref{sec:background:representation}) would only need to store the latter.

By switching from a vertex basis to an edge basis and aligning the Wedge Frontier with the layout of the destination-oriented edge data structure we are able to follow both Requirements 2 and 3.  Aligning itself with the destination-oriented edge data structure means that the Wedge Frontier is traversable using a destination orientation.  Furthermore, the Wedge Frontier operates at the granularity of edges rather than vertices, meaning that it enables filtering out individual edges within each destination vertex.  A pull-based graph processing engine is able to scan through the Wedge Frontier and limit message propagation to only those edges in the active subset of the graph, as these would be the only edges present in the Wedge Frontier.

Creation of the Wedge Frontier occurs by means of the transformation step shown in Figure~\ref{fig:wedge:system}, which consumes the traditional vertex-based source-oriented frontier as input.  The process is conceptually very simple: for each source vertex present in the traditional frontier produced by the pull engine, insert values into the Wedge Frontier that capture that vertex's out-edges.  In a hybrid framework we know each vertex's out-edges because they are encoded in an out-edge data structure like CSR.  The frontier transformation step consumes a similar kind of data structure, called an \textit{edge index}, except instead of encoding out-neighbors by vertex identifier it encodes out-edges by position within the in-edge data structure.  In other words, it maps each vertex to the Wedge Frontier values that capture its out-edges.  Therefore, we simply need to traverse this data structure and insert the values encoded within it into the Wedge Frontier.

Instantiating the edge index does not consume any additional memory over what a hybrid graph processing framework already consumes.  Because the source-oriented edge data structure used in a hybrid graph processing framework is not necessary in a pull-only framework, we can simply repurpose that space for the edge index.  In fact, doing so may actually decrease memory consumption because the source-oriented edge data structure might need to store edge weights, whereas the edge index does not.

Generating the Wedge Frontier represents a per-edge processing overhead: a sequential read to the edge index followed by a frontier insertion operation.  An iteration of an application executed using the pull engine along with the Wedge Frontier will outperform that same iteration executed using a push engine as long as the cost of a sequential access plus a frontier insertion operation is low enough so as not to overcome the throughput difference between push and pull.  A push engine performs a random-access atomic update operation to a large data structure (the vertex values) for each edge encountered, whereas the frontier transformation writes only to the frontier, which is much smaller and is therefore much more likely to result in cache hits when updated.  For more complicated applications that update multiple vertex values, a push engine would be burdened with heavier synchronization operations, such as per-vertex locks, whereas the frontier transformation overhead is fixed irrespective of application complexity.

\begin{figure}[t]
    \begin{center}
        \includegraphics{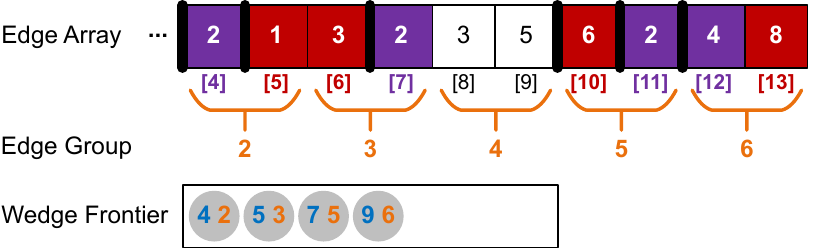}
    \end{center}
    \caption{Illustration of the frontier precision tuning parameter with 2 edges per group applied to the example in Figure~\ref{fig:wedge:frontierwedge}.  Instead of holding edge indices, the Wedge Frontier holds edge group numbers.  Red-highlighted edges are superfluous.}
    \label{fig:wedge:frontierprecision}
\end{figure}

\subsection{Frontier Transformation}
\label{sec:wedge:optimizations}

Executing the frontier transformation step can be quite expensive, particularly when the resulting Wedge Frontier is almost full, so blindly executing it every iteration would result in significantly slower performance than can be achieved using a hybrid framework.  The transformation executes in $O(|V| + |E|)$ time and is multi-threaded.  To make Wedge's frontier optimization design practical, we introduce two key optimizations, which we describe in this subsection.  Each optimization exposes a tuning parameter, and their respective impacts on performance are evaluated in \S\ref{sec:evaluation:tuning}.

\textbf{Frontier Fullness Threshold:}
To reduce the time spent in the frontier transformation step, we borrow the key concept of a hybrid framework and execute it selectively.  This is reflected in the frontier fullness decision shown in Figure~\ref{fig:wedge:system}.  If the pull engine produces as output a frontier that is sufficiently full, the following iteration is executed on the entire graph without a frontier, otherwise the transformation step produces the Wedge Frontier and the pull engine consumes it as input.  Whether a frontier is ``sufficiently full'' is determined by comparing the number of edges it contains to the \textit{frontier fullness threshold} tuning parameter.  Had we integrated the frontier transformation directly into the pull engine, we would have lost the ability to execute it selectively because metrics like frontier fullness cannot be evaluated until a processing iteration is fully completed.  For the applications and graphs we tested (\S\ref{sec:evaluation}), the ideal threshold value ranged from 1\% and 48\%, which resulted in up to 30\% of iterations running without the Wedge Frontier.

\textbf{Frontier Precision:}
Graphs often contain many more edges than vertices~\cite{ref_LAWS,ref_SNAP}, so switching the frontier from a vertex basis to an edge basis can incur two overheads.  First, an edge-based frontier will need to hold more values than a vertex-based frontier (one value per edge instead of one value per vertex), so the frontier data structure itself will consume more memory space.  Second, inserting vertices into an edge-based frontier takes longer than doing so into a vertex-based frontier because there are more values to insert.

To counteract the effects of these overheads, we introduce the \textit{frontier precision} tuning parameter, which can be adjusted to allow multiple edges to be represented using a single value in the Wedge Frontier.  The highest possible precision, whereby each edge is uniquely represented in the Wedge Frontier per Figure~\ref{fig:wedge:frontierwedge}, comes with the highest time and space overheads.  Lowering the frontier precision means grouping contiguous edges together; doing so can reduce the time and space overheads associated with the Wedge Frontier but at the expense of inserting a small number of extra edges into the Wedge Frontier that themselves are not part of the active subset.  For example, if we set the group size to 2 edges per group as depicted in Figure~\ref{fig:wedge:frontierprecision}, we see that the Wedge Frontier continues to represent the active subset of the graph but also includes some additional edges that the pull engine will process unnecessarily.  Edge group numbers identify edge groups by starting position in the edge topology data structure.  The maximum number of unnecessary edges is bounded by the group size and cannot grow arbitrarily large, thus avoiding the 1-million in-degree problem we identified in \S\ref{sec:wedge:design}.  We can arbitrarily reduce the frontier precision without introducing correctness issues because existing pull-based application implementations typically already support execution without any frontier.

\section{Implementation}
\label{sec:implementation}

Wedge is designed to be agnostic to specific implementation characteristics.  For the purposes of evaluation, it is prototyped in software and integrated into Grazelle~\cite{ref_Grazelle}, a state-of-the-art hybrid graph processing framework, resulting in a new pull-only version.  The purpose of this section is to describe the details of our software implementation of Wedge and its integration into Grazelle.  In particular, we address data structure selection, frontier transformation implementation, parallelization across multiple cores, and scaling to multiple processor sockets.  Grazelle uses Vector-Sparse to represent edge topology, so Wedge operates at the granularity of an edge vector such that the size of an edge group is measured in terms of the number of edge vectors, rather than individual edges, it contains.  Therefore, even at its highest precision, the Wedge Frontier does not uniquely identify each edge.  Wedge's frontier transformation step is implemented in approximately 140 LOC of C code, and modifications to Grazelle's pull engine total less than 50 LOC of C code.

{\bf Data Structures:}
The Wedge Frontier is implemented densely as a bit-mask, with one bit per edge group.  A sparse implementation is possible, following one of the main contributions in Ligra~\cite{ref_Ligra}, but we leave this as an engineering task for future work.  Grazelle also uses a dense bit-mask to represent its traditional frontiers, so mimicking this behavior in our implementation of Wedge facilitates a fairer comparison.  Furthermore, using a bit-mask comes with benefits such as fast insertion and automatic elimination of duplicates.  We experimented with implementing the Wedge Frontier using a hierarchical bit-mask data structure but observed no noticeable performance impact.  Using a single-level bit-mask is possible because Vector-Sparse does not depend on the vertex index to identify both ends of each edge (\S\ref{sec:background:representation}).  A Compressed-Sparse version would require one bit-mask for the vertex index and a second for the edge array.

The edge index follows the Compressed-Sparse-Row format.  However, instead of destination vertex identifiers, the values in the second-level edge array identify bit positions in the Wedge Frontier that need to be set for each source vertex identified in the first-level vertex index.  For each bit that is set in the traditional frontier, the frontier transformation step simply looks up that vertex in the first-level array of the edge index, iterates over its values in the second-level array of the edge index, and sets the corresponding bits in the Wedge Frontier.

{\bf Parallelization:}
Pull engine parallelization across multiple cores is unchanged from the manner in which Grazelle implements it.  We therefore refer interested readers to the original Grazelle publication for the details~\cite{ref_Grazelle}.  Parallelization of the Wedge transformation is implemented by slicing the traditional frontier into equally-sized pieces and dynamically scheduling each piece as threads become available to process them.  Pieces are sized statically; we consider neither the number of bits set to 1 within each piece nor the out-degrees of the vertices represented by each bit.  This decision has load balance implications, which we evaluate in \S\ref{sec:evaluation:scalability}.  We note, however, that load balance issues can be resolved using any known load balancing technique, such as work-stealing~\cite{ref_CilkPlus}, and leave this engineering task for future work.  In principle there is no need for synchronization between threads because all bit-setting operations are idempotent.  In practice, however, every such operation needs to be atomic because the addressable data unit is 1 byte, which gives rise to false sharing of bits within the same byte.

Scaling to multiple processor sockets occurs through graph partitioning, which is left to the graph processing framework.  Wedge assumes each socket has its own destination-oriented edge list, representing a partition of the graph, and locally generates a corresponding edge index for each one.  The traditional source-oriented vertex-based frontier is globally shared across sockets.  However, the Wedge Frontier is local, and one such data structure exists on each socket to correspond to the edge list partition for that same socket.  The frontier transformation runs locally and in parallel on each socket, consuming the global traditional frontier and producing the local Wedge Frontier.  The decision to run or skip frontier transformation is global, although in future work each socket could make this decision independently.

We use all available cores to perform the frontier transformation because, barring load balance issues, we found that increasing the number of cores continually increased performance (\S\ref{sec:evaluation:scalability}) and that even using all cores does not saturate the memory system.  In-memory compression, a topic studied in existing work~\cite{ref_LigraPlus}, can be applied to the edge index to reduce further the amount of memory bandwidth required and increase the potential performance to be gained by adding additional cores.  Furthermore, because it is application-independent and not limited by the memory system, the transformation step could conceivably be implemented as an accelerator.  If such an accelerator were built to share the processor's memory system, it would need to run the transformation step only when the pull engine is not running because the pull engine can saturate the memory system, and any memory bandwidth interference would reduce its performance.  Otherwise it can run in parallel with Grazelle as long as its operation could be terminated immediately and prematurely upon determination that production of the Wedge Frontier is unwarranted.

\section{Evaluation}
\label{sec:evaluation}

{\bf Server \& Datasets:}
We evaluate Wedge using a four-socket server equipped with four Intel Xeon E7-8890 v3 (18 physical/36 logical cores and 45\,MB LLC)~\cite{ref_IntelSpecLondon} processors and a total of 1\,TB DRAM running Ubuntu 14.04 LTS.  Our experiments use the six input datasets listed in Table~\ref{table:evaluation:datasets}.  All six are real-world datasets that span a wide variety of application areas and feature highly variable distributions of vertex degrees~\cite{ref_EvaluationStudy,ref_PerformanceStudy,ref_ExperimentalComparison}.  \texttt{dimacs-usa} and \texttt{twitter-2010} are often featured in the evaluations of other graph processing frameworks~\cite{ref_Ligra,ref_Polymer,ref_Galois,ref_X-Stream,ref_GraphChi}.  \texttt{dimacs-usa} is unique in that it is a mesh network, having a relatively small and even distribution of edges to vertices.  The others are scale-free graphs following a power-law degree distribution~\cite{ref_PowerGraph} of varying skew level.  The most extreme skew is found in \texttt{uk-2007}, which contains over $10\times$ more vertices having in-degree over 100,000 than \texttt{twitter-2010}, the second-most skewed graph~\cite{ref_WebGraph1,ref_WebGraph2,ref_Grazelle}.  Our plots sometimes refer to datasets by their shown abbreviations.

{\bf Applications:}
We focus our evaluation on three graph processing applications, all of which differ in their interaction with the frontier optimization: Breadth-First Search (BFS), Connected Components (CC), and Single-Source Shortest Path (SSSP).  In BFS, the frontier is initialized to contain just a single vertex (the root vertex of the traversal), and each vertex is inserted into the frontier for at most one iteration throughout the entire application.  The frontier begins extremely empty, grows in size, and finally empties fully, at which point the application converges.  Furthermore, because each vertex is only inserted into the frontier at most once, the frontier changes completely from one iteration to the next and generally remains very sparse.  CC is very much the opposite: the frontier is initialized to contain every vertex in the graph and gradually empties as the algorithm progresses.  Vertices are often inserted into the frontier for multiple iterations.  SSSP falls somewhere in the middle in that the frontier is initialized to contain a single vertex (the root vertex of the search), but the application behaves like CC in that there is no limit to the number of times a vertex may be inserted into the frontier.  SSSP is also the only of these applications that uses edge weights.  Edge weights do not affect frontier behavior but can affect the balance of performance between push and pull by biasing the memory access pattern towards sequential.  This is because edges are accessed sequentially in the edge list, and edge weights increase the amount of data that must be loaded per edge.  Our implementations of CC and SSSP are respectively based on HCC (label propagation \cite{ref_Pegasus}) and Bellman-Ford.  With the exception of \texttt{dimacs-usa}, the graphs listed in Table~\ref{table:evaluation:datasets} are unweighted, so for SSSP we generate weights using a multiplicative hash algorithm.

\begin{table}[t]
    \caption{Graph datasets used in experiments.}
    \label{table:evaluation:datasets}
    \begin{center}
        \begin{tabular}{|c|c|c|c|}
    \hline
    \textbf{Abbr.} &       \textbf{Name} &                                                 \textbf{Vertices} &     \textbf{Edges}          \\ \hline
    \hline
    \textbf{C} &           \texttt{cit-Patents}~\cite{ref_SNAP} &                          3.7M   &                16.5M                   \\ \hline
    \textbf{D} &           \texttt{dimacs-usa}~\cite{ref_DIMACS} &                         23.9M  &                58.3M                   \\ \hline
    \textbf{L} &           \texttt{livejournal}~\cite{ref_SNAP} &                          4.8M &                  69.0M                   \\ \hline
    \textbf{T} &           \texttt{twitter-2010}~\cite{ref_WebGraph2,ref_WebGraph1} &      41.7M  &                1.47B                   \\ \hline
    \textbf{F} &           \texttt{friendster}~\cite{ref_SNAP} &                           65.6M &                 1.81B                   \\ \hline
    \textbf{U} &           \texttt{uk-2007}~\cite{ref_WebGraph2,ref_WebGraph1} &           105.9M &                3.74B                   \\ \hline
\end{tabular}

    \end{center}
    \vspace{-0.2in}
\end{table}

We limit our evaluation scope to just these three applications because including other applications would not contribute any additional insights.  For example, PageRank and Collaborative Filtering are commonly evaluated in the literature~\cite{ref_Ligra,ref_X-Stream,ref_Polymer,ref_GraphX,ref_GraM,ref_GraphMat}, but they do not exploit the frontier optimization and are therefore irrelevant to our analysis.  Furthermore, other important applications are built on top of the fundamental applications we evaluate, meaning that anything learned from the studies we conduct carries over to those applications as well.  For example, implementations of Betweenness Centrality are based on BFS~\cite{ref_BC1,ref_BC2,ref_BC3}.

{\bf Experiments:}
\S\ref{sec:evaluation:wedge}, \S\ref{sec:evaluation:tuning}, and \S\ref{sec:evaluation:scalability} respectively provide in-depth analyses of performance characteristics, sensitivity to various tuning parameters, and scalability.  Experiments in \S\ref{sec:evaluation:wedge} and \S\ref{sec:evaluation:tuning} are executed using only a single socket, whereas multi-socket scaling experiments in \S\ref{sec:evaluation:scalability} use multiple.  Finally, in \S\ref{sec:evaluation:overall}, we present the overall performance of Wedge as compared to that of other state-of-the-art graph processing frameworks Grazelle~\cite{ref_Grazelle}, Ligra~\cite{ref_Ligra}, and GraphMat~\cite{ref_GraphMat}.  Both Ligra and Grazelle are hybrid graph processing frameworks that use the traditional source-oriented vertex-based frontier implementation.  Where they differ is that Ligra supports dynamically switching between sparse and dense representations of the frontier data structure, whereas Grazelle only supports a dense frontier representation but features much higher-throughput push-based and pull-based engines.  As a result, Ligra performs especially well for applications in which the active subset of the graph is consistently very small, such as BFS, outperforming Grazelle in some cases.  GraphMat is a push-only high-throughput graph processing framework built on a sparse matrix-vector multiplication back-end.  Prior to the introduction of Grazelle, GraphMat was considered the best-performing framework available.

\begin{figure*}[t]
    \begin{center}
        \begin{subfigure}[b]{0.99\linewidth}
            \begin{center}
                \includegraphics{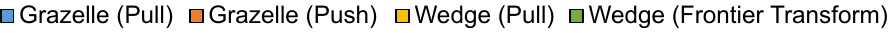}
            \end{center}
        \end{subfigure}
        \\
        \begin{subfigure}[b]{0.32\linewidth}
            \begin{center}
                \includegraphics{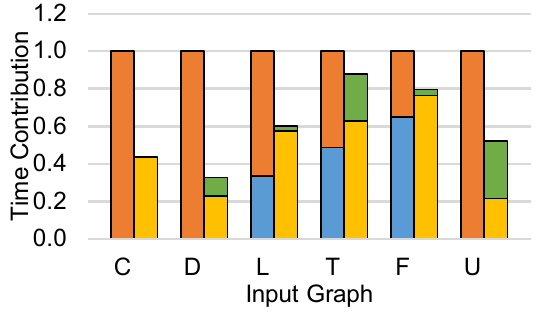}
            \end{center}
            \caption{Breadth-First Search}
            \label{fig:evaluation:wedge:bfs}
        \end{subfigure}
        \begin{subfigure}[b]{0.32\linewidth}
            \begin{center}
                \includegraphics{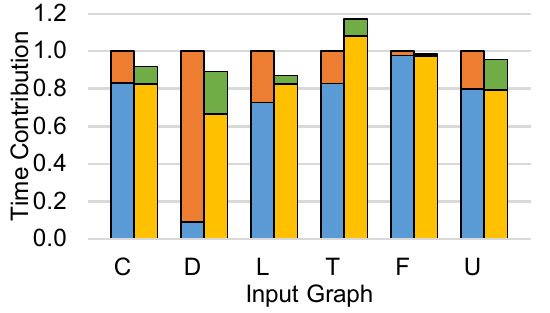}
            \end{center}
            \caption{Connected Components}
            \label{fig:evaluation:wedge:cc}
        \end{subfigure}
        \begin{subfigure}[b]{0.32\linewidth}
            \begin{center}
                \includegraphics{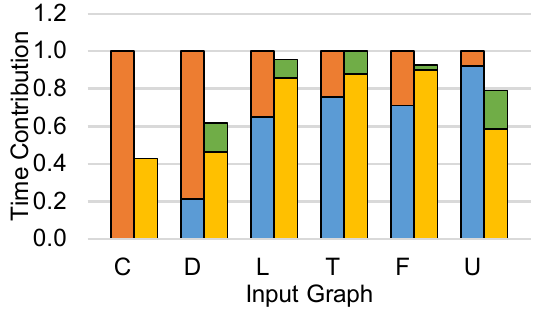}
            \end{center}
            \caption{Single-Source Shortest Path}
            \label{fig:evaluation:wedge:sssp}
        \end{subfigure}
    \end{center}
    \caption{Wedge performance relative to that of the Grazelle.  Results show the fraction of time spent in push and pull (Grazelle) or pull and frontier transformation (Wedge).}
    \label{fig:evaluation:wedge}
\end{figure*}

In all experiments except sensitivity tests (\S\ref{sec:evaluation:tuning}) we configure Wedge with experimentally-determined tuning parameters.  For CC and SSSP we use a frontier precision of 1 bit per 4 edge vectors and a frontier fullness threshold of 20\%.  For BFS we use a frontier precision of 1 bit per 8 edge vectors and a frontier fullness threshold of 1\%.  We use a much higher frontier fullness threshold for \texttt{uk-2007} because it is both extremely skewed (frontier size can grow very large) and has a high diameter (frontier size changes relatively slowly): 48\% for CC and SSSP, and 12\% for BFS.

\subsection{Wedge Performance}
\label{sec:evaluation:wedge}

Figure~\ref{fig:evaluation:wedge} provides an in-depth comparison of the execution time of Wedge and Grazelle.  Because Wedge is built on top of Grazelle, this comparison highlights the performance impact of switching from push to pull and the overheads of Wedge's frontier transformation step.  Also shown is the division of time between the two key parts of the execution in each case: push and pull for Grazelle and pull and frontier transform for Wedge.  Results are normalized to Grazelle's total runtime, shown as 1.0 in each plot.

The throughput improvements obtained by switching from push to pull are observable by comparing the height of the ``Wedge (Pull)'' bars to 1.0.  This is generally bound by the fraction of time Grazelle spends executing the push-based engine, which in turn is determined by the size of the active subset of the graph throughout the application.  Since BFS consistently maintains a relatively small active subset, the biggest difference is observed for BFS, followed by SSSP, and finally CC.  Most notably, Grazelle uses the push engine exclusively for BFS on \texttt{cit-Patents}, \texttt{dimacs-usa}, and \texttt{uk-2007}, so the throughput advantage of the pull pattern is maximally able to produce a performance improvement ($2.3\times$, $4.3\times$, and $4.8\times$, respectively).

Frontier transformation overheads are reflected in the ``Wedge (Frontier Transform)'' results and accounts for a relatively small percentage of the overall execution time.  Excessive time spent executing the Wedge transformation, marked by larger blocks on the plot (in particular BFS on \texttt{uk-2007}, which spends more time transforming the frontier than running the pull engine), is mostly attributable to issues of load balance between threads, which we discuss in more detail in \S\ref{sec:evaluation:scalability}.

An end-to-end performance comparison between Grazelle and Wedge can be made by comparing the overall bar heights between the two.  Performance improvement varies from approximately 1\% (CC executed on \texttt{friendster}) to $3\times$ (BFS executed on \texttt{dimacs-usa}).  In the case of CC executed on \texttt{twitter-2010} the pull engine's execution time with Wedge is slower than the overall execution time with Grazelle due to the added work that results from the imprecision of the Wedge Frontier.

It is clear from these results that Wedge matches or performs substantially better than Grazelle. Per Figure~\ref{fig:introduction:motivation}, a pull-only graph processing framework without Wedge's pull-friendly frontier optimization would be orders of magnitude slower than a hybrid framework on all three of these applications.  We therefore conclude that Wedge enables a pull-based engine to exploit the frontier optimization efficiently.

\subsection{Tuning Parameters}
\label{sec:evaluation:tuning}

\begin{figure*}[t]
    \begin{center}
        \begin{subfigure}[b]{0.99\linewidth}
            \begin{center}
                \includegraphics{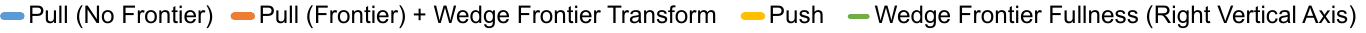}
            \end{center}
        \end{subfigure}
        \\
        \begin{subfigure}[b]{0.32\linewidth}
            \begin{center}
                \includegraphics{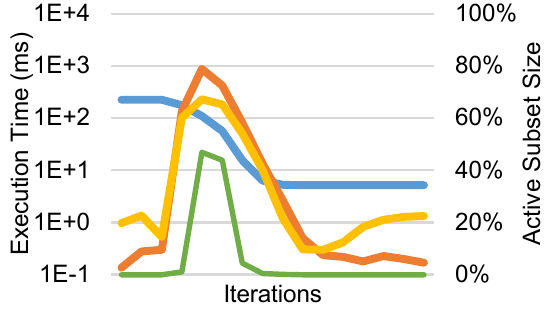}
            \end{center}
            \caption{BFS, twitter-2010}
            \label{fig:evaluation:tuning:threshold:bfs:t}
        \end{subfigure}
        \begin{subfigure}[b]{0.32\linewidth}
            \begin{center}
                \includegraphics{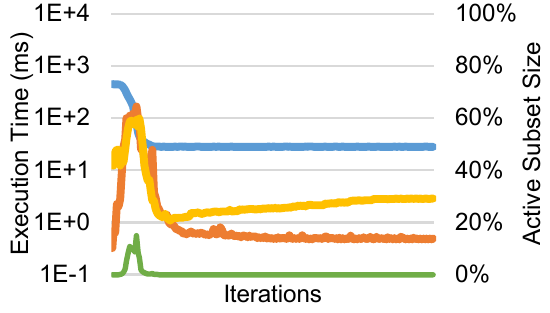}
            \end{center}
            \caption{BFS, uk-2007}
            \label{fig:evaluation:tuning:threshold:bfs:u}
        \end{subfigure}
        \begin{subfigure}[b]{0.32\linewidth}
            \begin{center}
                \includegraphics{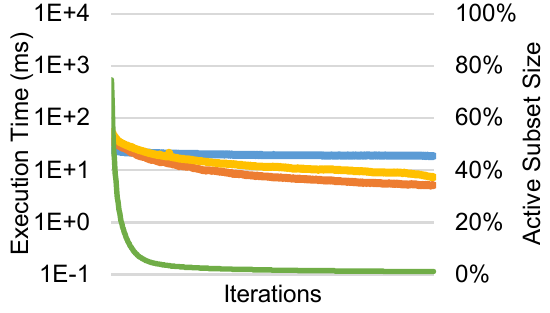}
            \end{center}
            \caption{CC, dimacs-usa}
            \label{fig:evaluation:tuning:threshold:cc:d}
        \end{subfigure}
        \\\vspace{0.05in}
        \begin{subfigure}[b]{0.32\linewidth}
            \begin{center}
                \includegraphics{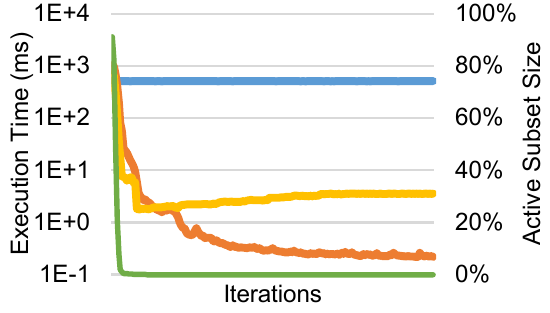}
            \end{center}
            \caption{CC, uk-2007}
            \label{fig:evaluation:tuning:threshold:cc:u}
        \end{subfigure}
        \begin{subfigure}[b]{0.32\linewidth}
            \begin{center}
                \includegraphics{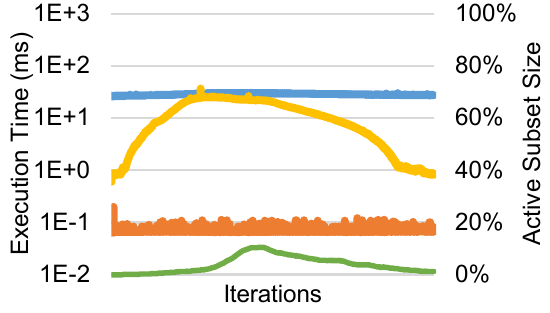}
            \end{center}
            \caption{SSSP, dimacs-usa}
            \label{fig:evaluation:tuning:threshold:sssp:d}
        \end{subfigure}
        \begin{subfigure}[b]{0.32\linewidth}
            \begin{center}
                \includegraphics{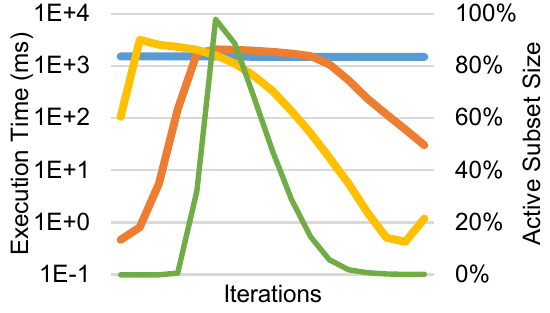}
            \end{center}
            \caption{SSSP, friendster}
            \label{fig:evaluation:tuning:threshold:sssp:f}
        \end{subfigure}
    \end{center}
    \caption{Iteration profiles for visualizing both the frontier fullness threshold and Wedge's benefit over push.  Horizontal axis is application iterations progressing in time from left to right.  Left vertical axis is logarithmic.}
    \label{fig:evaluation:tuning:threshold}
\end{figure*}

\begin{figure}[t]
    \begin{center}
        \includegraphics{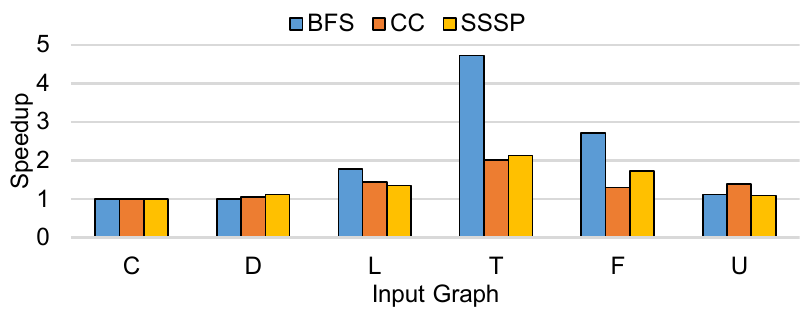}
    \end{center}
    \caption{Performance impact of the frontier fullness threshold optimization.  Baseline is the Wedge Frontier generated unconditionally.}
    \label{fig:evaluation:tuning:threshold:perf}
\end{figure}

{\bf Frontier Fullness Threshold:}
We evaluate the effect of frontier fullness threshold selection on Wedge performance by profiling the individual iterations of all three applications executed on each of the six graph datasets.  Our per-iteration profiles capture execution time without a Wedge Frontier (pull only, operating on the entire graph), execution time with a Wedge Frontier (pull plus Wedge transformation), execution time with the conventional push pattern, and the size of the active subset (percentage of edges in the Wedge Frontier).  The lattermost quantity is computed by summing the out-degrees of all the vertices the pull engine adds to the source-oriented vertex-based frontier it produces as output.  Normally the resulting value would be compared with the frontier fullness threshold to determine whether or not to produce a Wedge Frontier for the next iteration.  If less than the threshold a Wedge Frontier is generated, otherwise the pull engine runs without a Wedge Frontier.

Our plots, shown in Figure~\ref{fig:evaluation:tuning:threshold}, are intended to highlight the effectiveness of generating the Wedge Frontier selectively rather than unconditionally.  However, due to space limitations we cannot show every such combination and instead selected an assortment that highlights behavioral variety.  On the left vertical axis we show actual per-iteration execution times in milliseconds of the pull engine with and without a frontier, in the former case including the time taken by the frontier transformation step.  The goal in setting the frontier fullness threshold is to pick whichever mode produces a lower execution time for every iteration.  On the right vertical axis we show the size of the active subset, expressed as percentage of edges.  The horizontal axis shows iterations of the application progressing from left to right.  They are unnumbered because the numbers themselves are unimportant in this analysis.  Per-iteration results are also shown for Grazelle's push engine to illustrate the benefit of Wedge over using the push pattern.  Wedge generally outperforms the push engine when executed with a frontier, particularly at the tail ends of an application's execution.  Where it does not, differences are small enough and iteration times short enough as not to impact end-to-end performance noticeably.

Execution times of application iterations that use the Wedge Frontier scale with the size of the active subset because said size determines the amount of work in both the frontier transformation step and the accompanying pull engine iteration.  This is unlike the execution times of iterations that do not use the Wedge Frontier, which are relatively constant across all iterations.  Such iterations iterate over the entire graph and are therefore bound by pull engine throughput.  BFS is an exception and sees steadily decreasing no-frontier iteration execution times.  This is because vertices that have already been visited are skipped irrespective of the Wedge Frontier, and the number of visited vertices increases as the application progresses.

These results very clearly show a significant performance benefit to generating and consuming the Wedge Frontier when it is sufficiently empty.  Iterations at the left of Figure~\ref{fig:evaluation:tuning:threshold:sssp:f}, for example, are up to $3300\times$ faster when run with the Wedge Frontier than without.  Also apparent is the performance penalty associated with unconditionally generating the Wedge Frontier every iteration.  When the active subset is large enough, using the Wedge Frontier can result in a per-iteration slowdown of up to $8\times$, per Figure~\ref{fig:evaluation:tuning:threshold:bfs:t}.  This per-iteration performance difference can accumulate to the point of becoming dominant and resulting in reduced overall application performance.  End-to-end results (Figure~\ref{fig:evaluation:tuning:threshold:perf}) show that performance improves by up to almost $5\times$ by generating the Wedge Frontier only selectively.

\begin{figure}[t]
    \begin{center}
        \begin{subfigure}[b]{0.99\linewidth}
            \begin{center}
                \includegraphics{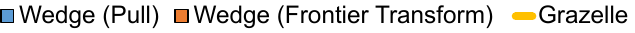}
            \end{center}
        \end{subfigure}
        \\
        \begin{subfigure}[b]{0.99\linewidth}
            \begin{center}
                \small{\textbf{\underline{Breadth-First Search}}}
            \end{center}
        \end{subfigure}
        \\\vspace{0.05in}
        \begin{subfigure}[b]{0.32\linewidth}
            \begin{center}
                \includegraphics{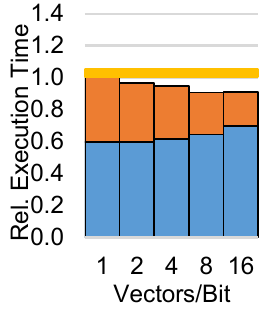}
            \end{center}
            \caption{twitter-2010}
            \label{fig:evaluation:tuning:precision:bfs:t}
        \end{subfigure}
        \begin{subfigure}[b]{0.32\linewidth}
            \begin{center}
                \includegraphics{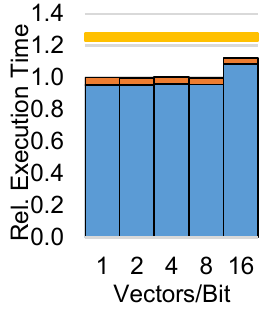}
            \end{center}
            \caption{friendster}
            \label{fig:evaluation:tuning:precision:bfs:f}
        \end{subfigure}
        \begin{subfigure}[b]{0.32\linewidth}
            \begin{center}
                \includegraphics{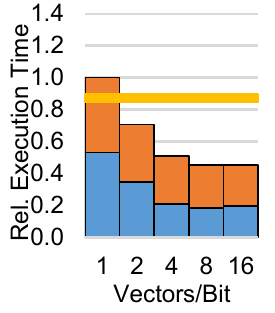}
            \end{center}
            \caption{uk-2007}
            \label{fig:evaluation:tuning:precision:bfs:u}
        \end{subfigure}
        \\\vspace{0.05in}
        \begin{subfigure}[b]{0.99\linewidth}
            \begin{center}
                \small{\textbf{\underline{Connected Components}}}
            \end{center}
        \end{subfigure}
        \\\vspace{0.05in}
        \begin{subfigure}[b]{0.32\linewidth}
            \begin{center}
                \includegraphics{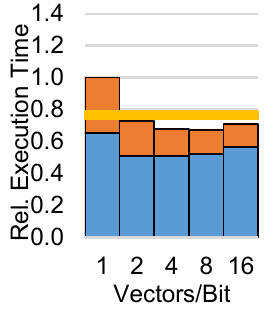}
            \end{center}
            \caption{dimacs-usa}
            \label{fig:evaluation:tuning:precision:cc:d}
        \end{subfigure}
        \begin{subfigure}[b]{0.32\linewidth}
            \begin{center}
                \includegraphics{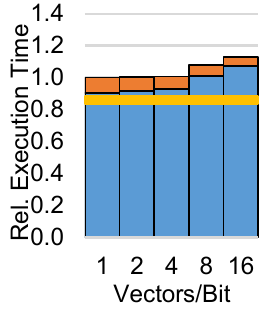}
            \end{center}
            \caption{twitter-2010}
            \label{fig:evaluation:tuning:precision:cc:t}
        \end{subfigure}
        \begin{subfigure}[b]{0.32\linewidth}
            \begin{center}
                \includegraphics{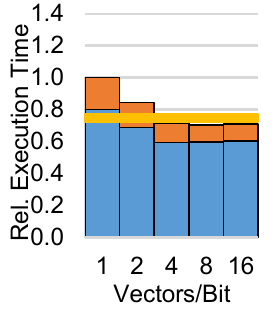}
            \end{center}
            \caption{uk-2007}
            \label{fig:evaluation:tuning:precision:cc:u}
        \end{subfigure}
        \\\vspace{0.05in}
        \begin{subfigure}[b]{0.99\linewidth}
            \begin{center}
                \small{\textbf{\underline{Single-Source Shortest Path}}}
            \end{center}
        \end{subfigure}
        \\\vspace{0.05in}
        \begin{subfigure}[b]{0.32\linewidth}
            \begin{center}
                \includegraphics{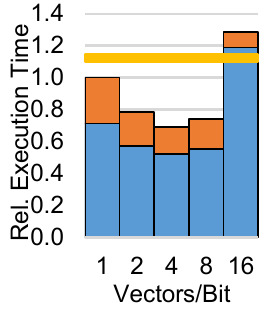}
            \end{center}
            \caption{dimacs-usa}
            \label{fig:evaluation:tuning:precision:sssp:d}
        \end{subfigure}
        \begin{subfigure}[b]{0.32\linewidth}
            \begin{center}
                \includegraphics{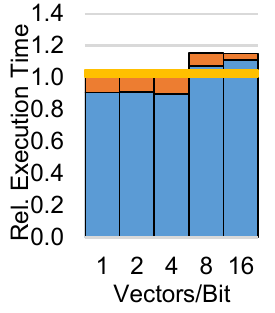}
            \end{center}
            \caption{twitter-2010}
            \label{fig:evaluation:tuning:precision:sssp:t}
        \end{subfigure}
        \begin{subfigure}[b]{0.32\linewidth}
            \begin{center}
                \includegraphics{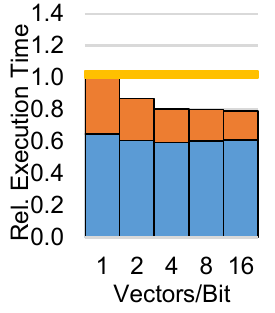}
            \end{center}
            \caption{uk-2007}
            \label{fig:evaluation:tuning:precision:sssp:u}
        \end{subfigure}
    \end{center}
    \caption{Sensitivity of Wedge performance to frontier precision.  Baseline for each plot is the total execution time at 1 vector per bit.}
    \label{fig:evaluation:tuning:precision}
\end{figure}

{\bf Frontier Precision:}
Figure~\ref{fig:evaluation:tuning:precision} shows performance sensitivity to frontier precision.  To conserve space we show results for a subset of the datasets such that various behaviors are captured.  Edge group size (edge vectors per Wedge Frontier bit) varies from 1 to 16.  Results are shown separately for the pull engine and the Wedge transformation step and are normalized to the overall execution time of the highest-precision configuration (1 bit per vector).  Normalized execution time of Grazelle is overlaid to provide context.

\begin{figure}[t]
    \begin{center}
        \begin{subfigure}[b]{0.99\linewidth}
            \begin{center}
                \includegraphics{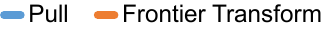}
            \end{center}
        \end{subfigure}
        \\
        \begin{subfigure}[b]{0.99\linewidth}
            \begin{center}
                \small{\textbf{\underline{Breadth-First Search}}}
            \end{center}
        \end{subfigure}
        \\\vspace{0.05in}
        \begin{subfigure}[b]{0.32\linewidth}
            \begin{center}
                \includegraphics{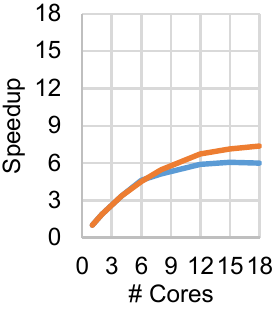}
            \end{center}
            \caption{dimacs-usa}
            \label{fig:evaluation:scalability:cores:performance:bfs:d}
        \end{subfigure}
        \begin{subfigure}[b]{0.32\linewidth}
            \begin{center}
                \includegraphics{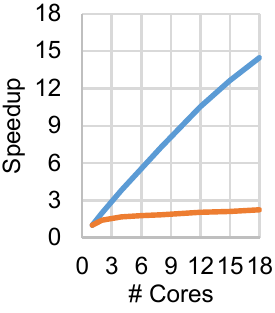}
            \end{center}
            \caption{twitter-2010}
            \label{fig:evaluation:scalability:cores:performance:bfs:t}
        \end{subfigure}
        \begin{subfigure}[b]{0.32\linewidth}
            \begin{center}
                \includegraphics{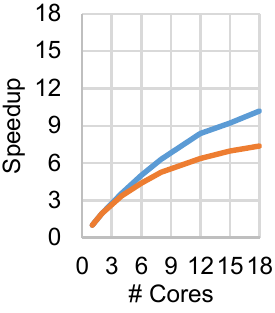}
            \end{center}
            \caption{uk-2007}
            \label{fig:evaluation:scalability:cores:performance:bfs:u}
        \end{subfigure}
        \\\vspace{0.05in}
        \begin{subfigure}[b]{0.99\linewidth}
            \begin{center}
                \small{\textbf{\underline{Connected Components}}}
            \end{center}
        \end{subfigure}
        \\\vspace{0.05in}
        \begin{subfigure}[b]{0.32\linewidth}
            \begin{center}
                \includegraphics{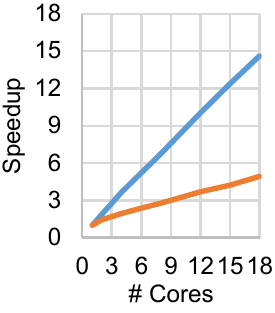}
            \end{center}
            \caption{livejournal}
            \label{fig:evaluation:scalability:cores:performance:cc:l}
        \end{subfigure}
        \begin{subfigure}[b]{0.32\linewidth}
            \begin{center}
                \includegraphics{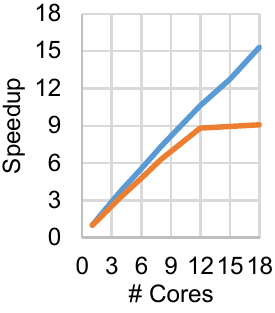}
            \end{center}
            \caption{twitter-2010}
            \label{fig:evaluation:scalability:cores:performance:cc:t}
        \end{subfigure}
        \begin{subfigure}[b]{0.32\linewidth}
            \begin{center}
                \includegraphics{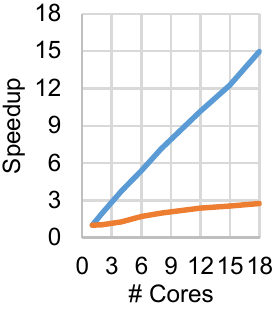}
            \end{center}
            \caption{friendster}
            \label{fig:evaluation:scalability:cores:performance:cc:f}
        \end{subfigure}
        \\\vspace{0.05in}
        \begin{subfigure}[b]{0.99\linewidth}
            \begin{center}
                \small{\textbf{\underline{Single-Source Shortest Path}}}
            \end{center}
        \end{subfigure}
        \\\vspace{0.05in}
        \begin{subfigure}[b]{0.32\linewidth}
            \begin{center}
                \includegraphics{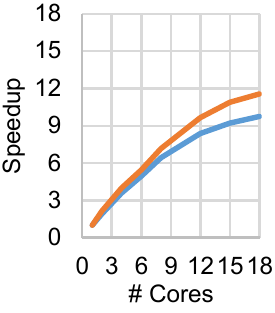}
            \end{center}
            \caption{dimacs-usa}
            \label{fig:evaluation:scalability:cores:performance:sssp:d}
        \end{subfigure}
        \begin{subfigure}[b]{0.32\linewidth}
            \begin{center}
                \includegraphics{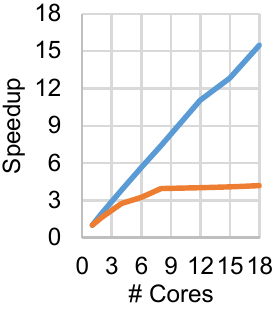}
            \end{center}
            \caption{twitter-2010}
            \label{fig:evaluation:scalability:cores:performance:sssp:t}
        \end{subfigure}
        \begin{subfigure}[b]{0.32\linewidth}
            \begin{center}
                \includegraphics{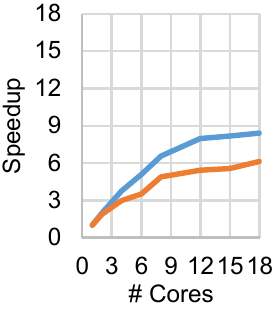}
            \end{center}
            \caption{uk-2007}
            \label{fig:evaluation:scalability:cores:performance:sssp:u}
        \end{subfigure}
    \end{center}
    \caption{Wedge single-socket multi-core performance scaling.  Baseline for each data point is the corresponding single-core performance.}
    \label{fig:evaluation:scalability:cores:performance}
\end{figure}

\begin{figure*}[t]
    \begin{center}
        \begin{subfigure}[b]{0.99\linewidth}
            \begin{center}
                \includegraphics{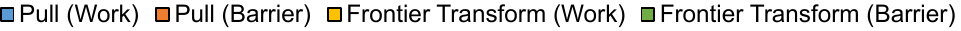}
            \end{center}
        \end{subfigure}
        \\
        \begin{subfigure}[b]{0.32\linewidth}
            \begin{center}
                \includegraphics{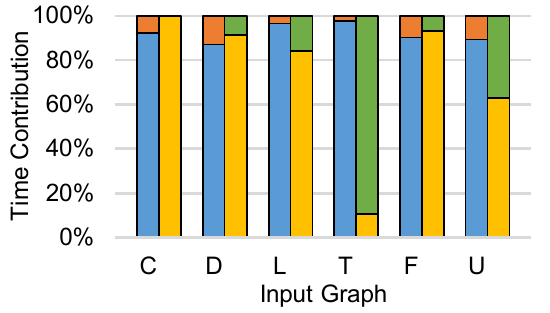}
            \end{center}
            \caption{Breadth-First Search}
            \label{fig:evaluation:scalability:cores:loadbalance:bfs}
        \end{subfigure}
        \begin{subfigure}[b]{0.32\linewidth}
            \begin{center}
                \includegraphics{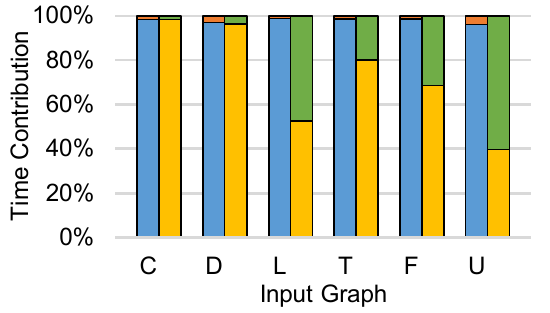}
            \end{center}
            \caption{Connected Components}
            \label{fig:evaluation:scalability:cores:loadbalance:cc}
        \end{subfigure}
        \begin{subfigure}[b]{0.32\linewidth}
            \begin{center}
                \includegraphics{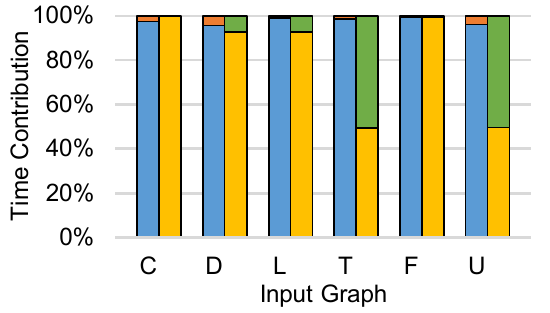}
            \end{center}
            \caption{Single-Source Shortest Path}
            \label{fig:evaluation:scalability:cores:loadbalance:sssp}
        \end{subfigure}
    \end{center}
    \caption{Wedge multi-core load balance effectiveness, expressed as average time division between processing and waiting at a synchronization barrier.  Lower barrier proportion is better.}
    \label{fig:evaluation:scalability:cores:loadbalance}
\end{figure*}

\begin{figure*}
    \begin{center}
        \begin{subfigure}[b]{0.99\linewidth}
            \begin{center}
                \includegraphics{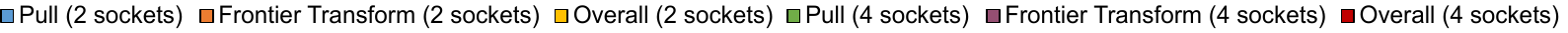}
            \end{center}
        \end{subfigure}
        \\
        \begin{subfigure}[b]{0.32\linewidth}
            \begin{center}
                \includegraphics{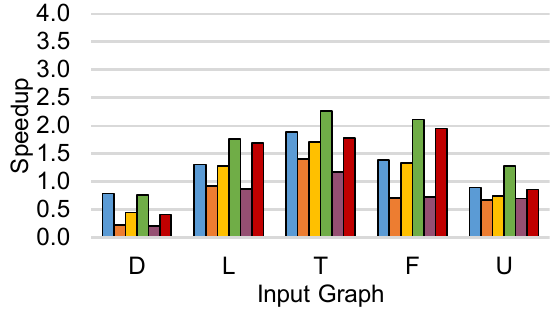}
            \end{center}
            \caption{Breadth-First Search}
            \label{fig:evaluation:scalability:sockets:bfs}
        \end{subfigure}
        \begin{subfigure}[b]{0.32\linewidth}
            \begin{center}
                \includegraphics{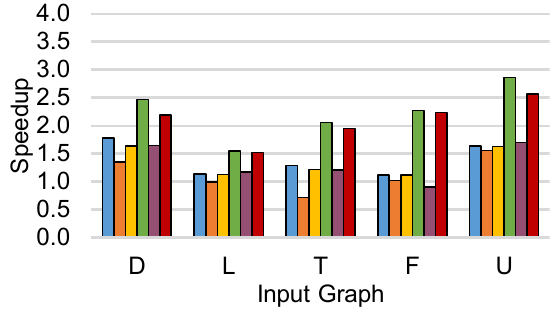}
            \end{center}
            \caption{Connected Components}
            \label{fig:evaluation:scalability:sockets:cc}
        \end{subfigure}
        \begin{subfigure}[b]{0.32\linewidth}
            \begin{center}
                \includegraphics{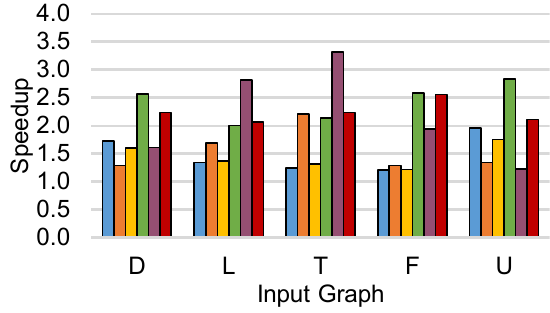}
            \end{center}
            \caption{Single-Source Shortest Path}
            \label{fig:evaluation:scalability:sockets:sssp}
        \end{subfigure}
    \end{center}
    \caption{Wedge multi-socket performance scaling, with all cores utilized on each socket.  Baseline for each data point is the corresponding single-socket performance.}
    \label{fig:evaluation:scalability:sockets}
\end{figure*}

In general we expect that a reduced frontier precision results in an increased pull engine time and a decreased frontier transformation time, trends we observe in many of the shown results.  The former is due to the increased number of unnecessary edges with which the pull engine is burdened: a single bit adds more edges to the active subset, which in turn increases the amount of potentially-unnecessary processing the pull engine must do.  The latter is due a reduction in the size of the Wedge Frontier data structure, which in turn produces two complementary effects.  First, a smaller Wedge Frontier improves caching efficacy.  Second, a greater number of edge vectors per bit reduces the number of bits that need to be set to capture the active subset, which reduces the number of memory write operations required.

Notable exceptions to these general trends are \texttt{dimacs-usa} and \texttt{uk-2007}, for which pull engine performance improves as frontier precision is reduced.  This behavior can occur when multiple contiguous Wedge Frontier bits are set in a higher-precision configuration such that reducing the precision does not introduce wasted work but rather reduces frontier-checking overheads.  For CC and SSSP, there is a second effect at play.  \texttt{dimacs-usa} and to a lesser extent \texttt{uk-2007} are characterized by having a relatively higher diameter than the other graph datasets in our evaluation.  This translates to the structures of these graphs having long chains of vertices along which messages must flow.  In a perfectly-precise frontier a message would propagate one hop per iteration.  With a less-precise frontier edges that would ordinarily not be processed until later iterations are processed earlier, leading to messages propagating multiple hops per iteration and significantly reducing the number of iterations required to reach convergence.  These effects have limits, evidenced by the slowdowns when transitioning from 8 to 16 edge vectors per bit for SSSP running on \texttt{dimacs-usa}.

Wedge uses a frontier precision of 4 or 8 edge vectors per bit depending on the application, the former for CC and SSSP and the latter for BFS.  This is generally a beneficial decision, resulting in a speedup of up to approximately $2\times$ and, at worst, a slowdown of 5\%.

\subsection{Scalability}
\label{sec:evaluation:scalability}

Performance scaling results within a single socket are shown in Figure~\ref{fig:evaluation:scalability:cores:performance}.  To conserve space we selected three representative cases per application, with the goal of showcasing differing behaviors.  Pull engine performance generally scales well with core count, limited primarily by saturation of the memory system.  This bottleneck, particularly visible in Figures~\ref{fig:evaluation:scalability:cores:performance:bfs:d}, \ref{fig:evaluation:scalability:cores:performance:sssp:d}, and ~\ref{fig:evaluation:scalability:cores:performance:sssp:u}, can be overcome by introducing locality optimizations~\cite{ref_Cagra} into the pull engine.

Performance of our implementation of the frontier transformation generally scales with increasing core count, but in many cases very slowly, limited by cores rather than by the memory system.  Scalability plateaus are primarily due to load imbalance between threads.  To quantify this we measured the time spent waiting at synchronization barriers for each thread and aggregated the results into Figure~\ref{fig:evaluation:scalability:cores:loadbalance}, which shows average per-thread time division between doing useful work (either in the pull engine or in the frontier transformation step) or being idle.  While the pull engine is effectively load-balanced, we observe that cases of limited scaling in Figure~\ref{fig:evaluation:scalability:cores:performance} are associated with up to 89\% of time wasted due to load imbalance.  This means that fixing our implementation, which can be done using any known software load balancing technique such as work-stealing~\cite{ref_CilkPlus}, could reduce time spent in the frontier transformation step by up to $9.1\times$.

Multi-socket performance scaling results are shown in Figure~\ref{fig:evaluation:scalability:sockets}.  Each data point is shown with respect to the corresponding single-socket result, and all results were obtained by fully utilizing the cores in each socket.  As a result, the theoretical maximum speedup is $2\times$ for two-socket results and $4\times$ for four-socket results.  Scalability of the pull engine is highly dependent on the proportion of node-local versus node-remote memory accesses, which in turn depends on the quality of the partitioning of the graph across sockets.  As Grazelle's partitioning scheme is very simple~\cite{ref_Grazelle}, it is not surprising that scaling results vary substantially between input datasets.  Nevertheless most results for CC and SSSP are between $2\times$ and $3\times$ with all four sockets active, indicating good scaling for these applications.  BFS is much more difficult to scale effectively because the active subset remains relatively small and is completely different each iteration.  The active subset is so small for \texttt{dimacs-usa} that many iterations slow down because the only active edges result in node-remote memory accesses.

Frontier transformation scaling characteristics differ from those of the pull engine.  Its workload is entirely dependent on the distribution of edges within the active subset, which affects load balance between sockets and is extremely difficult to predict.  Thus, even the best possible partitioning of the graph may be suboptimal for the transformation step, and in some cases the opposite is true.  Furthermore, node-remote memory accesses are guaranteed because the source-oriented vertex-based frontier produced by the pull engine must be consumed by all sockets running the frontier transformation step.  Nevertheless, for CC and SSSP it generally scales at least marginally with increased socket count, peaking at almost $3.5\times$ with all four sockets active.  As with the pull engine, BFS is difficult to scale and in some cases slows down.  Overall scaling results are biassed towards the scalability of the pull engine because the frontier transformation step does not dominate overall execution time (Figure~\ref{fig:evaluation:wedge}).

\subsection{Overall Comparison}
\label{sec:evaluation:overall}

\begin{figure*}[t]
    \begin{center}
        \begin{subfigure}[b]{0.19\linewidth}
            \begin{center}
                \includegraphics{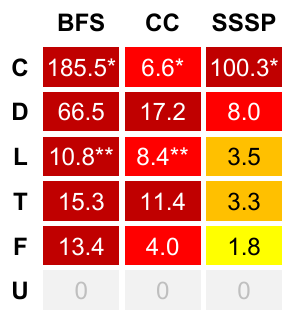}
            \end{center}
            \caption{GraphMat}
            \label{fig:evaluation:overall:graphmat}
        \end{subfigure}
        \begin{subfigure}[b]{0.19\linewidth}
            \begin{center}
                \includegraphics{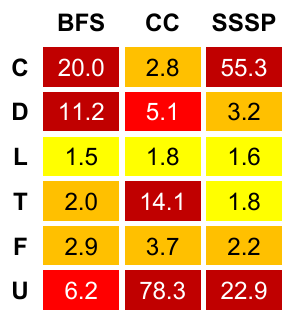}
            \end{center}
            \caption{Grazelle (Pull)}
            \label{fig:evaluation:overall:grazellepull}
        \end{subfigure}
        \begin{subfigure}[b]{0.19\linewidth}
            \begin{center}
                \includegraphics{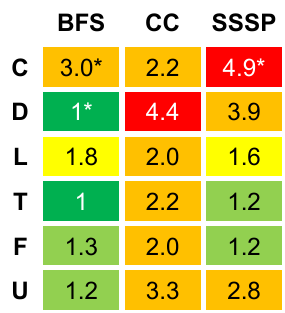}
            \end{center}
            \caption{Ligra}
            \label{fig:evaluation:overall:ligra}
        \end{subfigure}
        \begin{subfigure}[b]{0.19\linewidth}
            \begin{center}
                \includegraphics{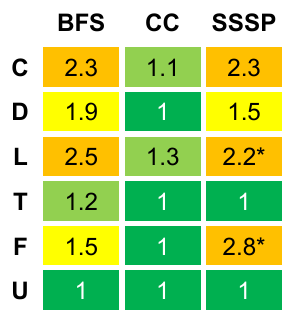}
            \end{center}
            \caption{Grazelle}
            \label{fig:evaluation:overall:grazelle}
        \end{subfigure}
        \begin{subfigure}[b]{0.19\linewidth}
            \begin{center}
                \includegraphics{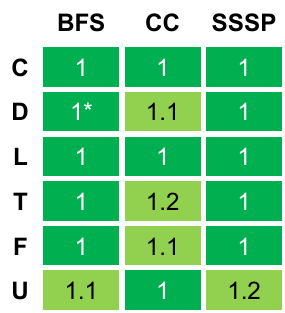}
            \end{center}
            \caption{Wedge}
            \label{fig:evaluation:overall:wedge}
        \end{subfigure}
    \end{center}
    \caption{Performance comparison between state-of-the-art frameworks, shown as heat maps with applications horizontal and graphs vertical.  Numbers display slowdown compared to best-observed execution time.  Default is a four-socket result; * and ** signify one-socket and two-socket results respectively.}
    \label{fig:evaluation:overall}
\end{figure*}

We compare the end-to-end performance of Wedge with that of Grazelle version 1.0.1~\cite{ref_GrazelleCode} configured both in hybrid mode and pull-only mode, Ligra version 1.5~\cite{ref_LigraCode}, and GraphMat version 1.0~\cite{ref_GraphMatCode}.  All frameworks were compiled with \texttt{gcc} version 5.5.0 and linked with whatever external libraries were recommended by the framework authors (Intel Cilk Plus~\cite{ref_CilkPlus} for Ligra, OpenMP~\cite{ref_OpenMP} for GraphMat).  Ligra allows the user to specify whether to compile it to use 32-bit or 64-bit integers internally.  Both Wedge and Grazelle use 64-bit integers to avoid artificially limiting graph size, so we compiled Ligra to do the same.  GraphMat only supports signed 32-bit integers internally and therefore cannot process \texttt{uk-2007} using any application because the number of edges exceeds their representational capability.

Results are shown as per-framework heat maps in Figure~\ref{fig:evaluation:overall} such that each number displays the slowdown compared to the fastest-observed time for that particular case.  Generally results were captured using all four sockets, but in some indicated cases it was faster to run with fewer sockets.  All cores in each socket are fully-utilized.

Wedge consistently outperforms Ligra, beating it by up to $3\times$ on BFS, up to $4\times$ on CC, and up to $4.9\times$ on SSSP.
It also generally outperforms Grazelle, though by a smaller margin (up to $2.8\times$) because both Grazelle and Wedge share pull engine implementations.  Its ability to outperform both demonstrates the effectiveness of Wedge at enabling a pull engine to execute a frontier-based application efficiently.  Part of the benefit of using Wedge over Ligra comes from its higher-throughput pull engine, but the remainder (and its entire benefit over Grazelle) is a result of replacing the time spent in the push engine with a smaller amount of time spent in the pull engine.  Because the active subset size with BFS is typically very small, Ligra's sparse frontier optimization is particularly effective, enabling it to outperform Grazelle in some cases.  However, using exclusively a pull engine with the Wedge Frontier is enough to more than overcome this performance gap.  Grazelle can outperform Wedge because the frontier transformation step scales less effectively with multiple sockets than Grazelle's push engine.  Single-socket results other than CC with \texttt{twitter-2010} favor Wedge.

GraphMat is uncompetitive with any of the other frameworks tested.  Despite being push-only and supporting the frontier optimization, its implementation of the frontier is suboptimal.  Rather than updating the frontier continuously as an iteration is executed, GraphMat compares old vertex values with new upon iteration completion and uses the results to create the frontier in a separate pass.

Arguably the most striking insight, and the key takeaway of this analysis, is obtained by comparing the performance of Grazelle running in pull-only mode (shown as ``Grazelle (Pull)'') with that of Wedge, essentially the ``before Wedge and after Wedge'' view of a pull engine.  Given that Grazelle's pull-only mode can be orders of magnitude slower than all other frameworks except GraphMat but Wedge's performance is almost always the best among the frameworks tested, we can conclude that Wedge has achieved our goal of rebuilding the frontier optimization such that a pull engine is able to exploit it efficiently.

\section{Related Work}
\label{sec:relatedwork}

Wedge is an entirely new approach that improves both the performance and utility of pull-based graph processing engines.  Existing work---whether it targets software, custom hardware, GPUs, or specialized accelerators---has predominantly focused on improving push-based engines~\cite{ref_X-Stream,ref_GraM,ref_Cagra,ref_Gunrock,ref_Graphicionado,ref_PowerGraph,ref_Chaos,ref_GraphXeonPhi}.  This is likely due to the conventional wisdom that pull-based engines cannot effectively exploit the frontier optimization, an idea that has persisted for years and represents the state-of-the-art~\cite{ref_PushPull}.  The pull pattern began as an optimization for the Breadth-First Search application specifically~\cite{ref_Beamer1,ref_Beamer2} but has since been generalized to other applications through the introduction of hybrid graph processing frameworks~\cite{ref_Grazelle,ref_Ligra,ref_Polymer}.  Garaph is the closest existing work to Wedge; its ``notify-pull'' approach proposes using a vertex-centric destination-oriented frontier that does not address the superfluous edge problem~\cite{ref_NotifyPull}.

Other areas of work in the graph processing community have improved graph processing performance through optimizations that target work scheduling across cores~\cite{ref_GraM,ref_Galois,ref_X-Stream}, graph partitioning across sockets~\cite{ref_GraM,ref_Polymer}, and optimizing synchronization and communication~\cite{ref_GraM,ref_Polymer,ref_Galois,ref_Grazelle}, which represent typical concerns for any parallel program~\cite{ref_ClassicParallel3,ref_ClassicParallel1,ref_ClassicParallel2}.  More recent work has attempted to take greater advantage of the underlying processor's hardware features, such as by improving caching effectiveness~\cite{ref_GraphLocality1,ref_GraphLocality2,ref_Cagra,ref_Locality}, reducing memory traffic through data structure compression~\cite{ref_LigraPlus}, and optimizing data structures for SIMD vectorization~\cite{ref_Grazelle}.  All of these strategies are orthogonal to Wedge and can generally be leveraged to improve the performance of both the frontier transformation step and the pull engine itself.

Domain-specific languages such as GraphIt~\cite{ref_GraphIt} combine many of these optimizations together into a single compiler.  They attempt to address the problem of requiring both push-based and pull-based versions of application code by generating both automatically from a single algorithm description.  However, even the state-of-the-art does not eliminate the need for multiple implementations: a second \textit{scheduling program} is required to assist the compiler in selecting appropriate optimizations.

\section{Conclusion}
\label{sec:conclusion}

Conventional wisdom states that a pull-based graph processing engine, despite having significantly higher throughput than a push-based engine, is fundamentally incapable of exploiting the frontier optimization.  Our key contribution in this work, \textit{Wedge}, defies this wisdom by rebuilding the frontier optimization so that it is suited for the pull pattern.

Wedge showcases a fundamentally new approach to addressing the trade-off between push and pull patterns in graph processing.  Prior work employed hybrid graph processing frameworks that support both patterns and dynamically switch between them.  This approach comes with two key disadvantages.  First, iterations that use push suffer from reduced performance.  Second, applications must be implemented multiple times (once each for push and pull).  Instead of continuing to juggle between both patterns, Wedge directly targets a pull engine's frontier-related deficiencies.

Wedge introduces the \textit{Wedge Frontier}, a representation of the frontier data structure that can efficiently be consumed by a pull engine, and a transformation step for generating it.  Because the transformation process can be expensive even when parallelized, we proposed two optimizations to make it practical.  First, we generate the Wedge Frontier only when it is sufficiently sparse.  Second, we coarsen the granularity of the representation to reduce both its size and the number of operations needed to generate it.

Wedge is implemented in software on top of Grazelle, a state-of-the-art hybrid graph processing framework, resulting in a new pull-only version.  It can outperform Grazelle, Ligra, and GraphMat respectively by up to $2.8\times$, $4.9\times$, and $185.5\times$.  Wedge's two key optimizations respectively improve its performance by up to $5\times$ and $2\times$.

\bibliographystyle{abbrv}
\balance
\bibliography{arxiv_wedge}

\end{document}